\documentclass[aps,prb,amsmath,amssymb,twocolumn,superscriptaddress]{revtex4-1}

\usepackage{graphicx}% Include figure files
\usepackage{dcolumn}% Align table columns on decimal point
\usepackage{bm}% bold math
\usepackage{hyperref}% add hypertext capabilities

\begin{document}

\title{Excitonic ordering in strongly correlated spin crossover systems: induced magnetism and excitonic excitation spectrum}

\author{Yu. S. Orlov}
\email{jso@iph.krasn.ru}
\affiliation{Siberian Federal University, Krasnoyarsk, 660041 Russia}
\affiliation{Kirensky Institute of Physics, Federal Research Center KSC SB RAS, Krasnoyarsk, 660036 Russia}

\author{S. V. Nikolaev}
\affiliation{Siberian Federal University, Krasnoyarsk, 660041 Russia}
\affiliation{Kirensky Institute of Physics, Federal Research Center KSC SB RAS, Krasnoyarsk, 660036 Russia}

\author{V. I. Kuz'min}
\affiliation{Kirensky Institute of Physics, Federal Research Center KSC SB RAS, Krasnoyarsk, 660036 Russia}

\author{A. E. Zarubin}
\affiliation{Kirensky Institute of Physics, Federal Research Center KSC SB RAS, Krasnoyarsk, 660036 Russia}

\author{S. G. Ovchinnikov}
\affiliation{Siberian Federal University, Krasnoyarsk, 660041 Russia}
\affiliation{Kirensky Institute of Physics, Federal Research Center KSC SB RAS, Krasnoyarsk, 660036 Russia}

\date{\today}

\begin{abstract}
The effects associated with interatomic hoppings of excitons and the excitonic Bose condensate formation in the strongly correlated spin crossover systems are considered in the framework of the effective Hamiltonian for the two-band Kanamori model. The appearance of antiferromagnetic ordering due to the exciton order is found even in the absence of interatomic exchange interaction. The spectrum of excitonic excitations is calculated at various points of the “temperature vs. crystal field” phase diagram. Outside the region of exciton ordering, the spectrum has a gap, which vanishes at the boundary of the exciton condensate phase. The non-uniform spectral weight distribution over the Brillouin zone is found. The role of electron-phonon interaction is discussed as well.

\end{abstract}

\maketitle

\section{\label{Intro}Introduction} 

The excitonic condensation and the excitonic insulator state have been under study for a long time, starting with the theoretical papers \cite{Mott, Knox, Kopaev}. As shown by Keldysh and Kopaev \cite{Kopaev}, the modified Bardeen-Cooper-Schrieffer (BCS) theory of superconductivity can be efficiently applied to describe metal-insulator phase transitions in semimetals. The Keldysh-Kopaev model of excitonic insulator has become the standard method of describing electronic correlations in the weak-interaction limit. In this model, a phase transition takes place at an arbitrarily weak interaction between electrons, which can be interpreted in analogy with a superconducting transition as a Bose condensation of weakly bound electron-hole pairs (large-radius excitons). In this paper, we consider the exciton condensate features formation, which is the condensation of local (at a crystal lattice site) magnetic excitons (small-radius excitons) in strongly correlated systems near the spin crossover.

Excitonic condensation in strongly correlated systems is under active discussion \cite{Kunes15, Nasu, Werner, Suzuki, Balents, Kaneko, Kunes14, Ikeda1, Platonov, Altarawneh, Rotter, Sotnikov, Tatsuno16, Ikeda2, Khaliullin}. An unexpected result of this work is the emergence of a long-range antiferromagnetic order due to excitonic ordering even in the absence of interatomic exchange interaction.

The spin crossover at zero temperature is a quantum phase transition occurring when varying pressure (crystal field). It is characterized by a topological order parameter defined by the geometrical Berry phase, which undergoes a step-like change by $2\pi$ at the transition point \cite{Nesterov}. Thus, it is of interest to study how quasiparticle (one-particle) and collective excitations change with spin crossover transition. In Ref.~\onlinecite{Orlov20} we showed that the electronic band structures in the low-spin (LS) and high-spin (HS) states are topologically nonequivalent and cannot be transformed into each other smoothly across a spin crossover transition. We are familiar with the paper \onlinecite{Nasu}, where, within the framework of the effective Hamiltonian obtained from the two-band Hubbard-Kanamori model, the spectrum of collective excitations in the exciton phase was studied in detail. In the present work, we will demonstrate the inhomogeneous spectral weight distribution of the exciton excitations over the Brillouin zone.

The results presented in this paper are obtained using the $X$-operator technique for the two-band Hubbard-Kanamori model. For a more detailed understanding of the obtained results, in Sec.~\ref{sec:5} an artificially simplified two-level model of one-electron states is introduced. It is defined by the Hamiltonian written in analogy with the Hamiltonian of the original problem in the fermion creation/annihilation operators representation. In the framework of such a model, the role of the electron-phonon interaction is discussed. It is found that, contrary to the diagonal electron-phonon interaction, the non-diagonal one leads to the opening of a gap in the excitonic excitation spectrum at the boundary of the excitonic condensate phase.

\section{\label{sec:2} Effective Hamiltonian}

A minimal model of strongly correlated spin crossover systems is the two-band Hubbard-Kanamori model. Its Hamiltonian can be written as 
\begin {equation}
H = H_\Delta   + H_t  + H_{Coulomb}.
\label {eq:1}
\end {equation}
Here, the first term
\begin {equation}
H_\Delta   = \varepsilon _1 \sum\limits_{i,\gamma } {a_{1i\gamma }^\dag  a_{1i\gamma }^{ } }  + \varepsilon _2 \sum\limits_{i,\gamma } {a_{2i\gamma }^\dag  a_{2i\gamma }^{ } } 
\label {eq:2}
\end {equation}
contains the one-ion energy of one-particle electron states with the energy levels $\varepsilon _1 $ and $\varepsilon _2  = \varepsilon _1  + \Delta $, where $\Delta$ is the crystal field energy (for convenience one can assume $\varepsilon _1 = 0$), $a_{\lambda i\gamma }^\dag$ creates a fermion at orbital $\lambda = 1,2$, site $i$, and with spin projection $\gamma = \pm 1/2$. The second term is
\begin {eqnarray}
H_t  &=& t_{11} \sum\limits_{\left\langle {i,j} \right\rangle ,\gamma } {a_{1i\gamma }^\dag  a_{1j\gamma }^{ } }  + t_{22} \sum\limits_{\left\langle {i,j} \right\rangle ,\gamma } {a_{2i\gamma }^\dag  a_{2j\gamma }^{ } } \nonumber \\
 &+& t_{12} \sum\limits_{\left\langle {i,j} \right\rangle ,\gamma } {\left( {a_{2i\gamma }^\dag  a_{1j\gamma }^{ }  + a_{1i\gamma }^\dag  a_{2j\gamma }^{ } } \right)},
\label {eq:3}
\end {eqnarray}
where $t_{\lambda \lambda '} $ is the nearest neighbor hopping parameter. The third term
\begin{widetext}
\begin{eqnarray}
 H_{Coulomb}  = U\sum\limits_{i,\lambda } {a_{i\lambda  \uparrow }^\dag  a_{i\lambda  \downarrow }^\dag  a_{i\lambda  \uparrow }^{ } a_{i\lambda  \downarrow }^{ } }  + V\sum\limits_{i,\lambda  \ne \lambda '} {a_{i\lambda  \uparrow }^\dag  a_{i\lambda ' \downarrow }^\dag  a_{i\lambda  \uparrow }^{ } a_{i\lambda ' \downarrow }^{ } } + V\sum\limits_{i,\lambda  > \lambda ',\gamma } {a_{i\lambda \gamma }^\dag  a_{i\lambda '\gamma }^\dag  a_{i\lambda \gamma }^{ } a_{i\lambda '\gamma }^{ } } \nonumber \\
+ J_H \sum\limits_{i,\lambda  > \lambda ',\gamma } {a_{i\lambda \gamma }^\dag  a_{i\lambda '\gamma }^\dag  a_{i\lambda '\gamma }^{ } a_{i\lambda \gamma }^{ } } + J_H \sum\limits_{i,\lambda  \ne \lambda '} {a_{i\lambda  \uparrow }^\dag  a_{i\lambda ' \downarrow }^\dag  a_{i\lambda ' \uparrow }^{ } a_{i\lambda  \downarrow }^{ } }  + J'_H \sum\limits_{i,\lambda  \ne \lambda '} {a_{i\lambda  \uparrow }^\dag  a_{i\lambda  \downarrow }^\dag  a_{i\lambda ' \uparrow }^{ } a_{i\lambda ' \downarrow }^{ } }
	\label {eq:4}
 \end{eqnarray}
\end{widetext}
includes the one-site Coulomb interaction energy (electron-electron interaction is considered in the Kanamori approximation with diagonal, in terms of orbital indices, matrix element $U$, nondiagonal $V$, and the Hund's exchange parameters $J_H$ and $J'_H$ \cite{Kanamori63}).

An important feature of such a two-orbital model is a possibility of formation, at half-filling ($N_e=2$ is an average number of electrons on a crystal lattice site) and in the zero hopping approximation $t_{\lambda\lambda'}=0$, of various localized two-electron states with spin values $S=0,1$, which makes possible a spin crossover with varying $\Delta$. Within the region $\Delta  < \Delta _C  = \sqrt {\left( {U - V + J_H} \right)^2  + {J_H'}^2 } $ the ground state is the triplet HS state ($S=1$) $\left| \sigma  \right\rangle$ with the energy $E_{HS}$:
\begin{eqnarray}
\left| \sigma  \right\rangle  = \left\{ \begin{array}{l}
 a_{1 \uparrow }^\dag  a_{2 \uparrow }^\dag  \left| 0 \right\rangle ,{\rm{   }}\sigma  =  + 1 \\ 
 \frac{1}{{\sqrt 2 }}\left( {a_{1 \uparrow }^\dag  a_{2 \downarrow }^\dag  \left| 0 \right\rangle  + a_{1 \downarrow }^\dag  a_{2 \uparrow }^\dag  \left| 0 \right\rangle } \right),{\rm{   }}\sigma  = 0\\ 
 a_{1 \downarrow }^\dag  a_{2 \downarrow }^\dag  \left| 0 \right\rangle ,{\rm{   }}\sigma  =  - 1, \\ 
 \end{array} \right.{\rm{   }} \nonumber
\end {eqnarray}
while at $\Delta  > \Delta _C $ the ground state is the singlet ($S=0$) state $\left| s \right\rangle  = C_1 \left( \Delta  \right)a_{1 \uparrow }^\dag  a_{1 \downarrow }^\dag  \left| 0 \right\rangle  - C_2 \left( \Delta  \right)a_{2 \uparrow }^\dag  a_{2 \downarrow }^\dag  \left| 0 \right\rangle $ with the energy $E_{LS} $, where $C_1 \left( \Delta  \right) = \sqrt {1 - C_2^2 \left( \Delta  \right)} $ and $C_2 \left( \Delta  \right) = {x \mathord{\left/ {\vphantom {x 2}} \right. \kern-\nulldelimiterspace} 2}\left( {1 + x + \sqrt {1 + x} } \right)$ are the normalizing coefficients \cite{Orlov21} ($x = {{{J'_H}^2 } \mathord{\left/
 {\vphantom {{{J'_H}^2 } {\Delta ^2 }}} \right.
 \kern-\nulldelimiterspace} {\Delta ^2 }}$).

To obtain an effective Hamiltonian it is convenient to use Hubbard $X$-operators $X^{p,q}  = \left| p \right\rangle \left\langle q \right|$ \cite{Hubbard} built on the eigenstates of the Hamilonian $H_\Delta   + H_{Coulomb}$:
 \begin{equation}
 \left( {H_\Delta + H_{Coulomb} } \right)\left| p \right\rangle  = E_p \left| p \right\rangle 
	\label {eq:5}
 \end{equation}
with the number of electrons taking values $N_e = 1,2,3$. Since the Hubbard operators form a linearly independent basis, any local operator can be represented as a linear combination of the $X$-operators, including the one-electron annihilation operator
\begin{equation}
a_{i\lambda \gamma }  = \sum\limits_{pq} {\left| p \right\rangle \left\langle {p\left| {a_{i\lambda \gamma } } \right|q} \right\rangle \left\langle q \right|}  = \sum\limits_{pq} {\chi _{\lambda \gamma } \left( {pq} \right)X_i^{pq} } 
\label {eq:6}
\end{equation}
Since the number of root vectors $\left( {pq} \right)$ \cite{Zaitsev} is finite, they can be enumerated by the local fermion quasiparticle number $m$ corresponding to each vector. Thus, $a_{i\lambda \gamma }  = \sum\limits_m {\chi _{\lambda \gamma } \left( m \right)X_i^m }$ and $a_{i\lambda \gamma }^\dag   = \sum\limits_m {\chi _{\lambda \gamma }^ *  \left( m \right)X_i^{\dag m} }$.

Using Eq.~\ref{eq:6}, the anomalous averages $\left\langle {a_{2f\gamma }^\dag  a_{1f\gamma }^{ } } \right\rangle $ without and with a spin projection change, $\left\langle {a_{2f\bar \gamma }^\dag  a_{1f\gamma }^{ } } \right\rangle $ ($\bar \gamma  =  - \gamma $), can be written as
\begin{equation}
\left\langle {a_{2f\gamma }^\dag  a_{1f\gamma }^{ } } \right\rangle  \approx  - \gamma \sqrt 2 \left( {C_2 \left\langle {X_f^{s,0} } \right\rangle  + C_1 \left\langle {X_f^{0,s} } \right\rangle } \right),
\label {eq:7}
\end{equation}
\begin{eqnarray}
&\left\langle {a_{2f\bar \gamma }^\dag  a_{1f\gamma } } \right\rangle&  \approx  - 2\gamma \left( {\gamma  + \frac{1}{2}} \right)\left( {C_2 \left\langle {X_f^{s, + 1} } \right\rangle  + C_1 \left\langle {X_f^{ + 1,s} } \right\rangle } \right) \nonumber \\
 &+& 2\gamma \left( {\gamma  - \frac{1}{2}} \right)\left( {C_2 \left\langle {X_f^{s, - 1} } \right\rangle  + C_1 \left\langle {X_f^{ - 1,s} } \right\rangle } \right).
\label {eq:8}
\end{eqnarray}

As follows from Eqs.~\ref{eq:7},~\ref{eq:8}, the exciton pairing is described by non-zero averages of singlet-triplet excitations. The Hamiltonian defined by Eq.~\ref{eq:1} can be rewritten in the $X$-operator representation as
\begin{equation}
H = \sum\limits_{i,p} {E_p X_i^{pp} }  + \sum\limits_{\left\langle {i,j} \right\rangle } {\sum\limits_{mn} {t^{mn} X_i^{\dag m} X_j^n } },
\label {eq:9}
\end{equation}
where $E_p $ is the multielectron eigenstate energy and $t^{mn}  = \sum\limits_{\lambda ,\lambda ',\gamma } {t_{\lambda \lambda '} \chi _{\lambda \gamma }^* \left( m \right)\chi _{\lambda '\gamma } \left( n \right)} $ is the renormalized hopping integral.

Using the Hamiltonian in Eq.~\ref{eq:9} as an initial one, we can obtain an effective Hamiltonian by excluding interband hopping. To do this, we apply the projection operator method developed in Ref.~\onlinecite{Chao} for the Hubbard model and in Ref.~\onlinecite{Gavrichkov} for the $p-d$ model (see also Refs.~\onlinecite{Kunes15, Nasu}). The obtained effective Hamiltonian is
\begin{equation}
H_{eff} = H_S + H_{nn} + H_{ex}.
\label {eq:10}
\end{equation}
Here, the first term describes an exchange contribution to the Heisenberg-like Hamiltonian, taking into account the energies of the electron confugurations in the LS and HS states:
\begin{equation}
H_S  = \frac{1}{2}J\sum\limits_{\left\langle {i,j} \right\rangle } {\left( { \mathbf{S}_i  \mathbf{S}_j  - \frac{1}{4} n_i n_j } \right)},
\label {eq:11}
\end{equation}
where $\mathbf{S}_i$ is the $S=1$ spin operator: $S_i^ +   = \sqrt 2 \left( {X_i^{ + 1,0}  + X_i^{0, - 1} } \right)$, $S_i^ -   = \sqrt 2 \left( {X_i^{0, + 1}  + X_i^{ - 1,0} } \right)$, and $S_i^z  = X_i^{ + 1, + 1}  + X_i^{ - 1, - 1} $ \cite{Valkov}; $J = {{\left( {t_{11}^2  + 2t_{12}^2  + t_{22}^2 } \right)} \mathord{\left/
 {\vphantom {{\left( {t_{11}^2  + 2t_{12}^2  + t_{22}^2 } \right)} {\Omega _g }}} \right.
 \kern-\nulldelimiterspace} {\Omega _g }}$ is the interatomic exchange interaction, $\Omega _g$ is the charge-transfer energy \cite{Chao, Gavrichkov}; $n_i  = 2\left( {X_i^{s,s}  + \sum\limits_\sigma  {X_i^{\sigma ,\sigma } } } \right) = 2\left( {n_i^{LS}  + n_i^{HS} } \right)$ is the particle number operator at site $i$ ($ n_i^{LS\left( {HS} \right)} $ is the occupation operator of the LS (HS) state). Using the completeness condition $X^{s,s}  + \sum\limits_\sigma  {X^{\sigma ,\sigma } }  = 1$, one can show that $\left\langle {n_i } \right\rangle  = 2\left( {\left\langle {n_i^{LS} } \right\rangle  + \left\langle {n_i^{HS} } \right\rangle } \right) = 2\left( {n_{LS}  + n_{HS} } \right) = 2$, where, here and below, angular brackets denote thermodynamical averages and $n_{LS\left( {HS} \right)} $ is an average LS (HS) occupation number ($n_{LS}  + n_{HS}  = 1$).

The next term,
\begin{equation}
H_{nn}  = \frac{1}{2}\tilde J\sum\limits_{\left\langle {i,j} \right\rangle } {X_i^{s,s}  \cdot X_j^{s,s} },
\label {eq:12}
\end{equation}
where $\tilde J = \left[ {1 - \left( {2C_1 C_2 } \right)^2 } \right]{{\left( {t_{11}^2  - 2t_{12}^2  + t_{22}^2 } \right)} \mathord{\left/
 {\vphantom {{\left( {t_{11}^2  - 2t_{12}^2  + t_{22}^2 } \right)} {\Omega _g }}} \right.
 \kern-\nulldelimiterspace} {\Omega _g }}$, represents a density-density type interaction of LS states.

The third term in Eq.~\ref{eq:10},
\begin{widetext}
\begin{equation}
H_{ex}  =  -\frac{{\varepsilon _S }}{2} \sum\limits_i \left({X_i^{s,s} }  - {\sum\limits_{\sigma  =  - S}^{ + S} {X_i^{\sigma ,\sigma } } } \right) + \sum\limits_\sigma  {\sum\limits_{\left\langle {i,j} \right\rangle } {\left[ {\frac{1}{2}J'_{ex} \left( {X_i^{\sigma ,s} X_j^{s,\sigma }  + X_i^{s,\sigma } X_j^{\sigma ,s} } \right) - \frac{1}{2}J''_{ex} ( - 1)^{\left| \sigma  \right|} \left( {X_i^{\sigma ,s} X_j^{\bar \sigma ,s}  + X_i^{s,\sigma } X_j^{s,\bar \sigma } } \right)} \right]} }, 
	\label {eq:13}
\end{equation}
\end{widetext}
contains interatomic hoppings of excitons with the amplitude $J'_{ex}$ as well as creation/annihilation processes of biexcitons on neighboring sites with the amplitude $J''_{ex}$. In the absence of any cooperative interactions, at negative values of the spin gap $\varepsilon _S  = E_{HS}  - E_{LS} $ the ground state is the HS state, whereas at positive spin gap values, the ground states is the LS state; $J'_{ex}  = 2C_1 C_2 {{\left( {t_{11} t_{22}  - t_{12}^2 } \right)} \mathord{\left/
 {\vphantom {{\left( {t_{11} t_{22}  - t_{12}^2 } \right)} {\Omega _g }}} \right.
 \kern-\nulldelimiterspace} {\Omega _g }}$, $J''_{ex}  = {{\left( {t_{11} t_{22}  - t_{12}^2 } \right)} \mathord{\left/
 {\vphantom {{\left( {t_{11} t_{22}  - t_{12}^2 } \right)} {\Omega _g }}} \right.
 \kern-\nulldelimiterspace} {\Omega _g }}$, $\bar \sigma  =  - \sigma$. The Hubbard operators $X_i^{\sigma ,s}$ and $X_i^{s,\sigma }$ in Eq.~\ref{eq:13} describe Bose-like excitations (excitons) between the LS singlet state $\left| s \right\rangle$ and the HS triplet state $\left| \sigma  \right\rangle$. The first term within the square brackets in Eq.~\ref{eq:13} describes the excitonic dispersion by means of interatomic hoppings (such a dispersion was considered long ago in the work of Vonsovskii and Svirskii \cite{Vonsovsky}). The second term in Eq.~\ref{eq:13} contains creation/annihilation processes of biexcitons at neighboring sites of a lattice, which makes the dispersion more complicated compared to the usual one obtained within the tight binding method \cite{Vonsovsky}. Near the spin crossover, the normalization constants defined above take values $C_1  \approx 1$ and $C_2  \approx 0$, thus, $J'_{ex}  \approx 0$ \cite{Orlov21}. At such circumstances, the biexciton excitations play the main role in the formation of the excitonic dispersion.

\section{\label{sec:3} Phase diagrams in the mean field approximation}
In mean field approximation (MF) for two sublattices $A$ and $B$, the terms in Eqs.~\ref{eq:11}--\ref{eq:13} can be expressed as the following Eqs.~\ref{eq:14}--\ref{eq:16}:

\begin{eqnarray}
H_S^{MF}  = zJm_B \sum\limits_{i_A } {S_{i_A }^z }  + zJm_A \sum\limits_{i_B } { S_{i_B }^z } \nonumber \\
 - zJ\frac{1}{4}n_B \sum\limits_{i_A } {n_{i_A } }  - zJ\frac{1}{4}n_A \sum\limits_{i_B } {n_{i_B } } \nonumber \\
  - \frac{1}{2}zJNm_A m_B  + \frac{1}{2}zJN,
\label {eq:14}
\end{eqnarray}
where $z$ is a number of nearest neighbors and $m_{A\left( B \right)}  = \left\langle {S_{i_{A\left( B \right)} }^z } \right\rangle$ is an $A$($B$)-sublattice magnetization;
\begin{eqnarray}
H_{nn}^{MF}  &=& z\tilde Jn_{LS,B} \sum\limits_{i_A } {n_{i_A }^{LS} }  + z\tilde Jn_{LS,A} \sum\limits_{i_B } {n_{i_B }^{LS} } \nonumber \\
&-& z\tilde J\frac{N}{2}n_{LS,A} n_{LS,B}.
	\label {eq:15}
\end{eqnarray}

The interaction proportional to $\tilde J$ leads to an additional cooperation mechanism, but in the following we will mainly neglect it to simplify the results, since it does not influence the behavior of phase diagrams qualitatively, only renormalizing the LS energies of the sublattices.

\begin{widetext}
\begin{eqnarray}
 H_{ex}^{MF}  = \sum\limits_F {\sum\limits_{\sigma  =  \pm 1,0} {\left\{ {zJ'_{ex} \Delta _{ex,\bar F}^\sigma  \sum\limits_{i_F } {\left( {X_{i_F }^{s,\sigma }  + X_{i_F }^{\sigma ,s} } \right)}  - \left( { - 1} \right)^{\left| \sigma  \right|} zJ''_{ex} \Delta _{ex,\bar F}^\sigma  \sum\limits_{i_F } {\left( {X_{i_F }^{s,\bar \sigma }  + X_{i_F }^{\bar \sigma ,s} } \right)} } \right.} } \nonumber  \\ 
 \left. { - \frac{1}{2}zN\left( {J'_{ex} \Delta _{ex,F}^\sigma  \Delta _{ex,\bar F}^\sigma   - \left( { - 1} \right)^{\left| \sigma  \right|} J''_{ex} \Delta _{ex,F}^\sigma  \Delta _{ex,\bar F}^{\bar \sigma } } \right)} \right\} - \varepsilon _S \sum\limits_{i_A } {X_{i_A }^{s,s} }  - \varepsilon _S \sum\limits_{i_B } {X_{i_B }^{s,s} }  + N\frac{{\varepsilon _S }}{2},
\label {eq:16}
\end{eqnarray}
\end{widetext}
where $F = \left(A,B\right)$ ($\bar F = A$ if $F = B$ and vice versa), $\Delta _{ex,A\left( B \right)}^\sigma   = \left\langle {X_{i_A \left( {i_B } \right)}^{s,\sigma } } \right\rangle $ are the excitonic order parameter components, which satisfy the equation $\left( {\Delta _{ex}^\sigma  } \right)^\dag   = \left\langle {X^{\sigma ,s} } \right\rangle  = \Delta _{ex}^\sigma $ at thermodynamic equilibrium. Note that, when $\Delta _{ex}^\sigma   \ne 0$, a quantum mechanical mixture of the LS and HS states is present, albeit in the absence of spin-orbital interaction. 

By solving the eigenstate problem
\begin{equation}
H_{eff}^{MF} \left| \psi  \right\rangle _k  = E_k \left| \psi  \right\rangle _k,
\label {eq:17}
\end{equation}
where $\left| \psi  \right\rangle _k  = C_{LS,k} \left| s \right\rangle  + \sum\limits_\sigma  {C_{{\rm{HS}},k} \left| \sigma  \right\rangle } $ are the eigenstates of the Hamiltonian $H_{eff}^{MF}  = H_S^{MF}  + H_{nn}^{MF}  + H_{ex}^{MF} $, and using the roots corresponding to the minimum of the free energy $F =  - k_B T\ln Z$, where $Z = \sum\limits_k {e^{ - {{E_k } \mathord{\left/
 {\vphantom {{E_k } {k_B T}}} \right. \kern-\nulldelimiterspace} {k_B T}}} } $ is the partition function, various thermodynamic averages included in $H_{eff}^{MF}$ can be calculated:
\begin{eqnarray}
\Delta _{ex,A\left( B \right)}^\sigma   = \sum\limits_k {\frac{{\left\langle {\psi _k \left| {X_{i_A \left( {i_B } \right)}^{s,\sigma } } \right|\psi _k } \right\rangle e^{{{ - E_k } \mathord{\left/
 {\vphantom {{ - E_k } {k_B T}}} \right.
 \kern-\nulldelimiterspace} {k_B T}}} }}{Z}}  \nonumber \\
m_{A\left( B \right)}  = \sum\limits_k {\frac{{\left\langle {\psi _k \left| {S_{i_{A\left( B \right)} }^z } \right|\psi _k } \right\rangle e^{{{ - E_k } \mathord{\left/
 {\vphantom {{ - E_k } {k_B T}}} \right.
 \kern-\nulldelimiterspace} {k_B T}}} }}{Z}} \nonumber \\
n_{HS,A\left( B \right)}  = \sum\limits_k {\frac{{\left\langle {\psi _k \left| {\sum\limits_\sigma  {X_i^{\sigma ,\sigma } } } \right|\psi _k } \right\rangle e^{{{ - E_k } \mathord{\left/
 {\vphantom {{ - E_k } {k_B T}}} \right.
 \kern-\nulldelimiterspace} {k_B T}}} }}{Z}}\nonumber.
\end{eqnarray}
Thus, when solving Eq.~\ref{eq:17}, one deals with a self-consistent problem of finding the eigenstates and the eigenvalues of the effective Hamiltonian in the mean field approximation.

\begin{figure*}
\includegraphics[width=17.2cm]{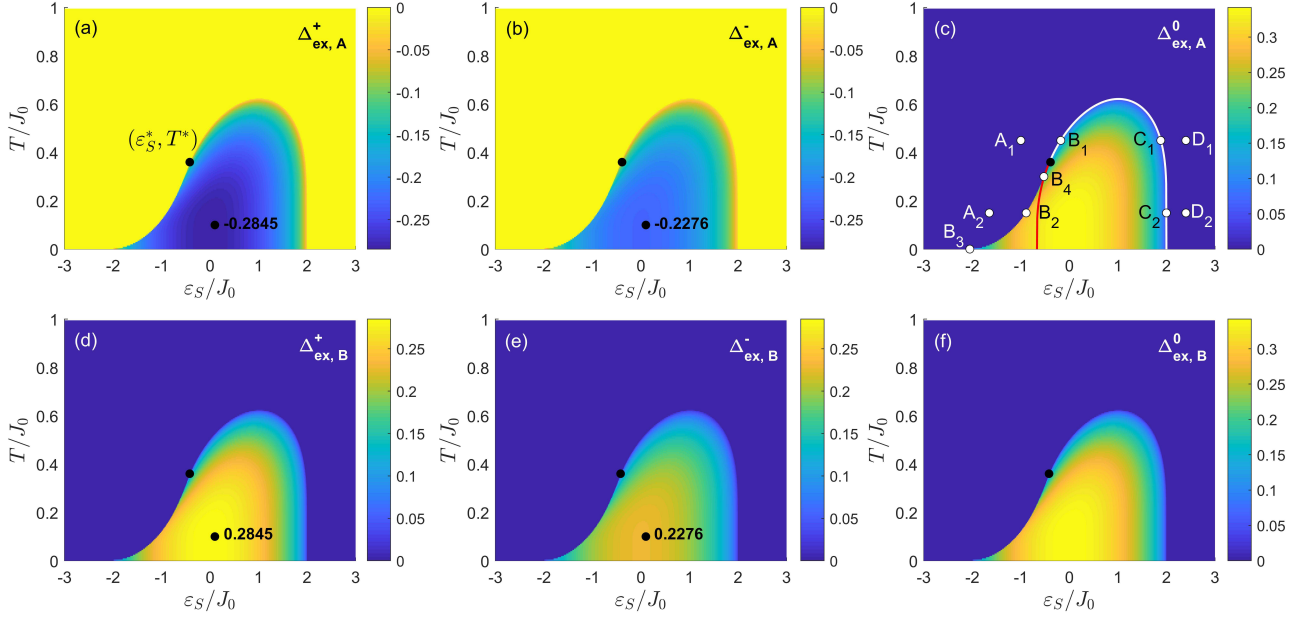}
\caption{\label{fig:1} The calculated phase diagrams of the excitonic order parameter components $\Delta _{ex}^\sigma  $ for the sublattices $A$ and $B$; the parameters are $z = 4$, $J''_{ex}  = 0.5 J_0 $, $J_0  = 28 \text{K}$.}
\end{figure*}

\begin{figure}
\includegraphics[width=8.6cm]{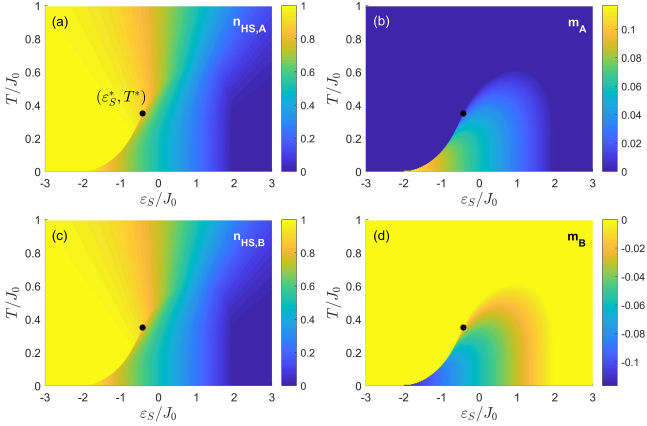}
\caption{\label{fig:2} The phase diagrams of (a), (c) the HS state occupation and (b), (d) magnetization for the both sublattices.}
\end{figure}

Figures~\ref{fig:1} and~\ref{fig:2} show the dependence of the excitonic order parameter components $\Delta _{ex}^\sigma$, the HS state occupation $n_{HS} $, and magnetization $m$ for the two sublattices $A$ and $B$ on temperature $T$ and spin gap (crystal field) $\varepsilon _S $. The calculations were done omitting the interatomic exchange ($J=0$). However, to compare conveniently with the $J\neq0$ case \cite{Orlov21}, $T$ and $\varepsilon _S $ are shown in units of the exchange integral $J = J_0  = 28\text{K}$ \cite{Hoch}; $z = 4$, $J''_{ex}  = 0.5J_0 $. From Fig.~\ref{fig:2} (a), (c) it is seen that $n_{HS,A}  = n_{HS,B} $;  $m_A  =  - m_B $, so the long-range antiferromagnetic ordering is realized (Fig.~\ref{fig:2} (b), (d)) even at $J=0$, since $\Delta _{ex,A\left( B \right)}^ +   \ne \Delta _{ex,A\left( B \right)}^ -  $ (Fig.~\ref{fig:1} (a), (b) and (d), (e)), $\Delta _{ex,A}^0  = \Delta _{ex,B}^0 $ (Fig.~\ref{fig:1} (c), (f)), and $\Delta _{ex,A}^{ + / - } = -\Delta _{ex,B}^{ + / - } $; for example, in Fig.~\ref{fig:1} the values of $\Delta _{ex,A\left( B \right)}^\sigma  $ at the point ($\varepsilon _S / J_0  = 0.1$, $T / J_0 = 0.1$ ) are shown.

The phase diagrams in Figs.~\ref{fig:1} and \ref{fig:2} clearly show an existence of a special point, which is the tricritical point $\left(T^*, \varepsilon^*_S \right)$, where the line of the second order phase transitions smoothly transforms into the line of first order phase transitions. In the region $\varepsilon _S  > \varepsilon _S^ *$ (see Fig.~\ref{fig:2} (b), (d)) the system undergoes a second order phase transition from an antiferromagnetic (HS) state to a paramagnetic state; contrarily, at $\varepsilon _S  < \varepsilon _S^*$ there is a first order phase transition.

We note that all the presented phase diagrams are asymmetrical with respect to the change of the sign of the spin gap. Contrarily, the toy model considered below in Sec.~\ref{sec:5} possesses such a symmetry. The difference in the multielectron terms (the phase diagrams in Figs~\ref{fig:1} and ~\ref{fig:2} are obtained using the two-electron singlet and triplet terms) is related to the different degeneracy multiplicity of the HS and LS states, which leads to a broken symmetry with respect to the spin gap sign inversion.

\section{\label{sec:4} Spectrum of excitons}
Let us consider the two-particle Green functions in terms of the initial fermion operators to describe collective (in terms of the electron system) excitonic excitations without a spin projection change 

\begin{equation}
{\mathbf{G}}_{\left( 2 \right)}  = \left( {\begin{array}{*{20}c}
   {\left\langle {\left\langle {{c_{1f\gamma }^\dag  c_{2f\gamma } }}
 \mathrel{\left | {\vphantom {{c_{1f\gamma }^\dag  c_{2f\gamma } } {c_{2g\gamma }^\dag  c_{1g\gamma } }}}
 \right. \kern-\nulldelimiterspace}
 {{c_{2g\gamma }^\dag  c_{1g\gamma } }} \right\rangle } \right\rangle _\omega  } & {\left\langle {\left\langle {{c_{1f\gamma }^\dag  c_{2f\gamma } }}
 \mathrel{\left | {\vphantom {{c_{1f\gamma }^\dag  c_{2f\gamma } } {c_{1g\gamma }^\dag  c_{2g\gamma } }}}
 \right. \kern-\nulldelimiterspace}
 {{c_{1g\gamma }^\dag  c_{2g\gamma } }} \right\rangle } \right\rangle _\omega  }  \\
   {\left\langle {\left\langle {{c_{2f\gamma }^\dag  c_{1f\gamma } }}
 \mathrel{\left | {\vphantom {{c_{2f\gamma }^\dag  c_{1f\gamma } } {c_{2g\gamma }^\dag  c_{1g\gamma } }}}
 \right. \kern-\nulldelimiterspace}
 {{c_{2g\gamma }^\dag  c_{1g\gamma } }} \right\rangle } \right\rangle _\omega  } & {\left\langle {\left\langle {{c_{2f\gamma }^\dag  c_{1f\gamma } }}
 \mathrel{\left | {\vphantom {{c_{2f\gamma }^\dag  c_{1f\gamma } } {c_{1g\gamma }^\dag  c_{2g\gamma } }}}
 \right. \kern-\nulldelimiterspace}
 {{c_{1g\gamma }^\dag  c_{2g\gamma } }} \right\rangle } \right\rangle _\omega  }
\end{array}} \right)
\label {eq:18}
\end{equation}
and with a spin projection change

\begin{equation}
\bar {\mathbf{G}}_{\left( 2 \right)}^ \pm   = \left( {\begin{array}{*{20}c}
   {\left\langle {\left\langle {{c_{1f\gamma }^\dag  c_{2f\bar \gamma } }}
 \mathrel{\left | {\vphantom {{c_{1f\gamma }^\dag  c_{2f\bar \gamma } } {c_{2g\bar \gamma }^\dag  c_{1g\gamma } }}}
 \right. \kern-\nulldelimiterspace}
 {{c_{2g\bar \gamma }^\dag  c_{1g\gamma } }} \right\rangle } \right\rangle _\omega  } & {\left\langle {\left\langle {{c_{1f\gamma }^\dag  c_{2f\bar \gamma } }}
 \mathrel{\left | {\vphantom {{c_{1f\gamma }^\dag  c_{2f\bar \gamma } } {c_{1g\bar \gamma }^\dag  c_{2g\gamma } }}}
 \right. \kern-\nulldelimiterspace}
 {{c_{1g\bar \gamma }^\dag  c_{2g\gamma } }} \right\rangle } \right\rangle _\omega  }  \\
   {\left\langle {\left\langle {{c_{2f\gamma }^\dag  c_{1f\bar \gamma } }}
 \mathrel{\left | {\vphantom {{c_{2f\gamma }^\dag  c_{1f\bar \gamma } } {c_{2g\bar \gamma }^\dag  c_{1g\gamma } }}}
 \right. \kern-\nulldelimiterspace}
 {{c_{2g\bar \gamma }^\dag  c_{1g\gamma } }} \right\rangle } \right\rangle _\omega  } & {\left\langle {\left\langle {{c_{2f\gamma }^\dag  c_{1f\bar \gamma } }}
 \mathrel{\left | {\vphantom {{c_{2f\gamma }^\dag  c_{1f\bar \gamma } } {c_{1g\bar \gamma }^\dag  c_{2g\gamma } }}}
 \right. \kern-\nulldelimiterspace}
 {{c_{1g\bar \gamma }^\dag  c_{2g\gamma } }} \right\rangle } \right\rangle _\omega  }
\end{array}} \right)
\label {eq:19}
\end{equation}

Using Eq.~\ref{eq:6}, one can write Eqs.~\ref{eq:18},~\ref{eq:19} as
\begin{equation}
\mathbf{G}_{\left( 2 \right)}  \approx \frac{1}{2}\left( {C_2^2  - C_1^2 } \right)\left( {\begin{array}{*{20}c}
   {G_{fg}^0 \left( \omega  \right)} & { - L_{fg}^0 \left( \omega  \right)}  \\
   {L_{fg}^0 \left( \omega  \right)} & { - G_{fg}^0 \left( \omega  \right)}  \\
\end{array}} \right)
\label {eq:20}
\end{equation}
and
\begin{equation}
\bar {\mathbf{G}}_{\left( 2 \right)}^ \pm   \approx \left( {C_1^2  - C_2^2 } \right)\left( {\begin{array}{*{20}c}
   {G_{fg}^ \pm  \left( \omega  \right)} & { - L_{fg}^ \pm  \left( \omega  \right)}  \\
   {L_{fg}^ \pm  \left( \omega  \right)} & { - G_{fg}^ \pm  \left( \omega  \right)}  \\
\end{array}} \right),
\label {eq:21}
\end{equation}
where
\begin{equation}
G_{fg}^\sigma  \left( \omega  \right) = \left\langle {\left\langle {{X_f^{s,\sigma } }}
 \mathrel{\left | {\vphantom {{X_f^{s,\sigma } } {X_g^{\sigma ,s} }}}
 \right. \kern-\nulldelimiterspace}
 {{X_g^{\sigma ,s} }} \right\rangle } \right\rangle _\omega,
\label {eq:22}
\end{equation}
\begin{equation}
L_{fg}^\sigma  \left( \omega  \right) = \left\langle {\left\langle {{X_f^{\bar \sigma ,s} }}
 \mathrel{\left | {\vphantom {{X_f^{\bar \sigma ,s} } {X_g^{\sigma ,s} }}}
 \right. \kern-\nulldelimiterspace}
 {{X_g^{\sigma ,s} }} \right\rangle } \right\rangle _\omega.
\label {eq:23}
\end{equation}
The ``+(-)'' sign in Eq.~\ref{eq:21} corresponds to the $\gamma = \uparrow \left(\downarrow\right)$ spin projection in Eq.~\ref{eq:19}.

Within Hubbard-1 approximation $\left\langle {X_f^{\sigma ' \ne \sigma ,\sigma } } \right\rangle  = 0$. In the two-sublattice case one obtains
\begin{eqnarray}
&G&_{AA,\mathbf{k}}^\sigma  \left( \omega  \right) = F_{A,\sigma } \left[  - J'_{ex} \left( \mathbf{k} \right)G_{AB,\mathbf{k}}^\sigma  \left( \omega  \right) \right. \nonumber \\
&+& \left. ( - 1)^{\left| \sigma  \right|} J''_{ex} \left( \mathbf{k} \right)L_{AB,\mathbf{k}}^\sigma  \left( \omega  \right) -  1 \right]  /  \left(\omega  -  \varepsilon_B \right),
\label {eq:24}
\end{eqnarray}
\begin{eqnarray}
\label {eq:25}
&G&_{AB,\mathbf{k}}^\sigma  \left( \omega  \right) = F_{B,\sigma } \left[  - J'_{ex} \left( \mathbf{k} \right)G_{AA,\mathbf{k}}^\sigma  \left( \omega  \right)  \right. \nonumber \\
&+& \left. ( - 1)^{\left| \sigma  \right|} J''_{ex} \left( \mathbf{k} \right)L_{AA,\mathbf{k}}^\sigma  \left( \omega  \right) \right] / \left(\omega  - \varepsilon_A \right),
\end{eqnarray}
\begin{eqnarray}
&L&_{AA,\mathbf{k}}^\sigma  \left( \omega  \right) = F_{A,\sigma } \left[ J'_{ex} \left( \mathbf{k} \right)L_{AB,\mathbf{k}}^\sigma  \left( \omega  \right) \right. \nonumber \\
&-& \left. ( - 1)^{\left| \sigma  \right|} J''_{ex} \left( \mathbf{k} \right)G_{AB,\mathbf{k}}^\sigma  \left( \omega  \right) \right]/ \left(\omega  + \varepsilon_B \right),
\label {eq:26}
\end{eqnarray}
\begin{eqnarray}
&L&_{AB,\mathbf{k}}^\sigma  \left( \omega  \right) = F_{B,\sigma } \left[ J'_{ex} \left( \mathbf{k} \right)L_{AA,\mathbf{k}}^\sigma  \left( \omega  \right) \right. \nonumber \\
&-& \left. ( - 1)^{\left| \sigma  \right|} J''_{ex} \left( \mathbf{k} \right)G_{AA,\mathbf{k}}^\sigma  \left( \omega  \right) \right] / \left(\omega  + \varepsilon_A \right),
\label {eq:27}
\end{eqnarray}
where $\varepsilon _{A\left(B\right)} = \varepsilon _S  + \sigma \frac{1}{2}zJm_{A\left(B\right)}$, $F_{A\left( B \right),\sigma }  = \left\langle {X_{A\left( B \right)}^{\sigma ,\sigma } } \right\rangle  - \left\langle {X_{A\left( B \right)}^{s,s} } \right\rangle $; $J'_{ex} \left( \mathbf{k} \right)$ and  $J''_{ex} \left( \mathbf{k} \right)$ are the Fourier transforms of $J'_{ex} $ and $J''_{ex} $. Within the region of the exciton ordering, the occupation numbers and the factors $F_{A\left( B \right),\sigma } $ depend on projection $\sigma$ and subblatice numbers. Thus, magnetization is induced at even $J=0$, as it is shown in Fig.~\ref{fig:2} (b),(d). 

From Eqs.~\ref{eq:24}--\ref{eq:27}, the following excitonic spectrum can be obtained:
\begin{eqnarray}
\omega _{\mathbf{k},\sigma }^2  = \varepsilon _S^2  + F_{A,\sigma } F_{B,\sigma } \left( {{J'_{ex}} ^2 \left( \mathbf{k} \right) - {J''_{ex}} ^2 \left( \mathbf{k} \right)} \right) \nonumber \\
\pm 2J'_{ex} \left( \mathbf{k} \right)\varepsilon _S \sqrt {F_{A,\sigma } F_{B,\sigma } } 
\label {eq:28}
\end{eqnarray}

The excitonic order parameter does not appear explicitly into Eqs.~\ref{eq:24}--\ref{eq:27} defining the Green functions. However, it is related to the occupation numbers $\left\langle {X_{A\left( B \right)}^{\sigma ,\sigma } } \right\rangle $ and $\left\langle {X_{A\left( B \right)}^{s,s} } \right\rangle $ calculated within the self-consistent problem given by Eq.~\ref{eq:17}. Outside the excitonic region (when $\Delta _{ex,A\left( B \right)}^\sigma   = 0$), one has $F_{A\left( B \right),\sigma }  = F$ (see Fig.~\ref{fig:1} (c),(f)). This way, at $J=0$, instead of Eq.~\ref{eq:28}, one can use the expression
\begin{equation}
\omega _{\mathbf{k},\sigma }^ \pm   =  \pm \sqrt {\left( {\varepsilon _S  - FJ'_{ex} \left( \mathbf{k} \right)} \right)}^2  - F^2 {J''_{ex}} ^2 \left( \mathbf{k} \right).
\label{eq:29}
\end{equation}

\begin{figure}[h]
\includegraphics[width=8.6cm]{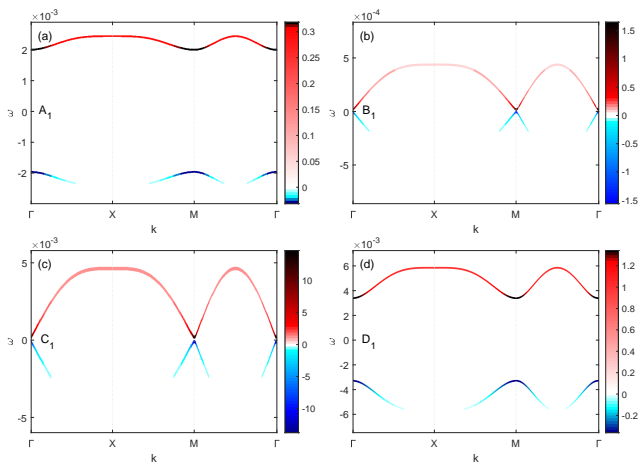}
\caption{\label{fig:3} The excitonic spectrum $\omega _{\mathbf{k},\sigma }^ \pm  $ defined by Eq.~\ref{eq:29} at the points $A_1$, $B_1$, $C_1$, and $D_1$ (along the $T /J_0  = 0.45$ - line) of the phase diagram in Fig.~\ref{fig:1} (c). Here and below the color is proportional to the spectral weight of the excitations.}
\end{figure}

\begin{figure}[h]
\includegraphics[width=8.6cm]{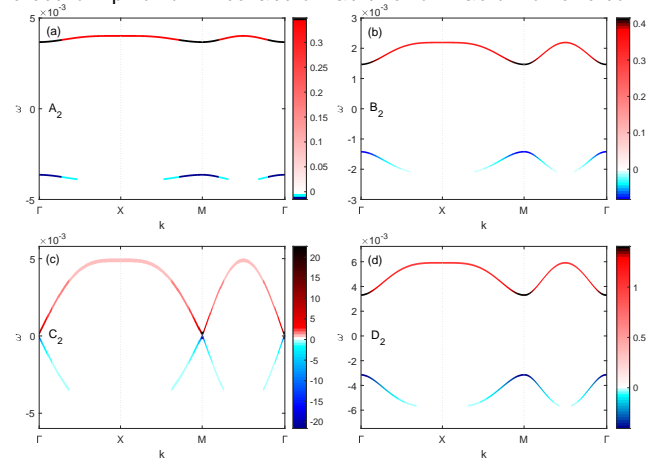}
\caption{\label{fig:4} The excitonic spectrum $\omega _{\mathbf{k},\sigma }^ \pm  $ defined by Eq.~\ref{eq:29} at the points $A_2$, $B_2$, $C_2$, and $D_2$ (along the $T /J_0  = 0.15$ - line) of the phase diagram in Fig.~\ref{fig:1} (c).}
\end{figure}

\begin{figure*}
\includegraphics[width=17.2cm]{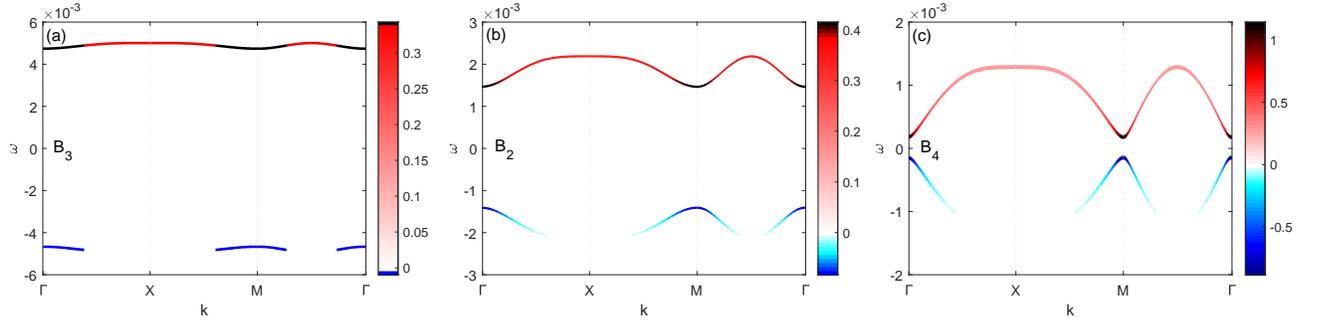}
\caption{\label{fig:5} The excitonic spectrum $\omega _{\mathbf{k},\sigma }^ \pm  $ defined by Eq.~\ref{eq:29} at the points $B_3$, $B_2$, and $B_4$, (along the boundary of the second order phase transition) of the phase diagram in Fig.~\ref{fig:1} (c).}
\end{figure*}

In Figs.~\ref{fig:3}--\ref{fig:5}, the spectrum $\omega _{\mathbf{k},\sigma }^ \pm $ calculated at different points labeled as $A$, $B$, $C$, and $D$ in Fig.~\ref{fig:1} (c), is drawn in the case of a 2D square lattice within the paramagnetic Brillouin zone. Here and below $\Gamma \left( {0,0} \right)$, $\text{X}\left( {\pi,0} \right)$, and $\text{M}\left( {\pi,\pi} \right)$ are the high-symmetry points.
In Fig.~\ref{fig:1} (c), the white (before the tricritical point) and red (after the tricritical point) line marks the boundary of the region within which the spectrum defined by Eq.~\ref{eq:28} becomes complex, which means that the normal state of the system is unstable with respect to the formation of the excitonic condensate. The white curve strictly coincides with the second order phase transition line. The red curve indicates what the boundary of the normal state instability would look like if the region of the first order phase transition did not exist. However, in our calculations, this curve enters the excitonic region right after the tricritical point ($\varepsilon _S^ *  ,{\rm{ }}T^ *  $) and strictly coincides with the right boundary of the metastable state region, which always exist in the case of a first order phase transition.

The upper band has a positive spectral weight, which points at ``quasiparticle'' excitations, whereas the lower band is of a ``hole'' type and has negative spectral weight (Fig.~\ref{fig:3}--\ref{fig:5}). When these bands merge at $\omega=0$, a transition to a new state associated with the formation of the excitonic condensate occurs - the whole spectral weight concentrates at the points $\Gamma$ and $\text{M}$ (Fig.~\ref{fig:3}--\ref{fig:5}). The spectral weight is distributed non-uniformly among the Brillouin zone. Particularly, it is suppressed around the points $\left( {\pi,0} \right)$ and $\left( \pi/2, \pi/2 \right)$ in the lower band at any value of $\varepsilon _S $. Concerning the Fermi-type quasiparticle excitations, their unusial spectral weight distribution (inversion) due to topological properties of spin crossover was discussed in our work \onlinecite{Orlov20}, where the electronic band structure of the strongly correlated spin crossover systems was considered both in the LS and HS states.

Figure~\ref{fig:5} shows that the gap in the spectrum remains non-zero throughout the first order phase transition line, and the value of the gap decreases while approaching the tricritical point, at which it vanishes.


\section{\label{sec:5} Toy model}

The Hamiltonian in Eq.~\ref{eq:16} has a significantly complex phase diagram and a multicomponent excitonic oder parameter. Let us try to simplify as much as possible the case considered above and analyze the influence of electron-phonon interaction at the formation of excitonic condensate using an example of an artificially simplified Hamiltonian of a two-level system with local one-electron states ``1'' and ``2'' with the energies $\varepsilon_1 $ and $\varepsilon_2 $ ($\varepsilon = \varepsilon_2 - \varepsilon_1$). Instead of Eq.~\ref{eq:13} we will now use the analogous in essence Eq.~\ref{eq:30} in terms of one-particle fermionic operators  $c_{\lambda i\gamma }$ ($c_{\lambda 1\gamma }^\dag  $) of annihilation (creation) of electrons at site $i$, state $\lambda=1,2$ with spin projection $\gamma  =  \pm 1/2$:

\begin{widetext}
\begin{eqnarray}
 \hat H_{ex}  &=& \varepsilon _1 \sum\limits_{i,\gamma } {c_{1i\gamma }^\dag  c_{1i\gamma }^{ } }  + \varepsilon _2 \sum\limits_{i,\gamma } {c_{2i\gamma }^\dag  c_{2i\gamma }^{ } } + \frac{{J'_1 }}{2}\sum\limits_{\left\langle {i,j} \right\rangle ,\gamma } {\left( {c_{1i\gamma }^\dag  c_{2i\gamma }^{ } c_{2j\gamma }^\dag  c_{1j\gamma }^{ }  + h.c.} \right)} + \frac{{J'_2 }}{2}\sum\limits_{\left\langle {i,j} \right\rangle ,\gamma } {\left( {c_{1i\gamma }^\dag  c_{2i\bar \gamma }^{ } c_{2j\bar \gamma }^\dag  c_{1j\gamma }^{ }  + h.c.} \right)}   \nonumber \\
&+& \frac{{J''_1 }}{2}\sum\limits_{\left\langle {i,j} \right\rangle ,\gamma } {\left( {c_{2i\gamma }^\dag  c_{1i\gamma }^{ } c_{2j\gamma }^\dag  c_{1j\gamma }^{ }  + h.c.} \right)}  + \frac{{J''_2 }}{2}\sum\limits_{\left\langle {i,j} \right\rangle ,\gamma } {\left( {c_{2i\gamma }^\dag  c_{1i\bar \gamma }^{ } c_{2j\bar \gamma }^\dag  c_{1j\gamma }^{ }  + h.c.} \right)} 
\label{eq:30}
\end{eqnarray}
\end{widetext}
For example, whereas in Eq.~\ref{eq:13} the operator $X_i^{\sigma ,s} $ describes a transition from the two-particle singlet state $\left| s \right\rangle $ to the triplet state $\left| \sigma  \right\rangle $, an analogous role in Eq.~\ref{eq:30} is played by the operator structure $c_{2j\gamma }^\dag  c_{1j\gamma }^{ } $ (without a change of a spin projection) or $c_{2j\bar \gamma }^\dag  c_{1j\gamma }^{ } $ (with a change of spin projection). Below we will drop the latter: $J'_2  = J''_2  = 0$
, so $J'_1  = J'_{ex} $ and $J''_1  = J''_{ex} $,

Taking into account the electron-phonon interaction one has
\begin{equation}
H = H_{ex} + H_{1ph} + H_{2ph},
\label{eq:31}
\end{equation}
where
\begin{widetext}
\begin{equation}
 H_{1ph}  = \hbar {\omega _0} _{\left( 1 \right)} \sum\limits_i {\left( {a_i^\dag  a_i^{ }  + \frac{1}{2}} \right)}  - \frac{1}{2}V_a \sum\limits_{\left\langle {i,j} \right\rangle } {\left( {a_i^{ }  + a_i^\dag  } \right)\left( {a_j^{ }  + a_j^\dag  } \right)}  + g_1 \sum\limits_{i,\gamma } {\left( {a_i^{ }  + a_i^\dag  } \right)\left( {c_{1i\gamma }^\dag  c_{1i\gamma }^{ }  - c_{2i\gamma }^\dag  c_{2i\gamma }^{ } } \right)}
\label{eq:32}
\end{equation}
\end{widetext}
contains the diagonal electron-phonon interaction. Next, the term
\begin{widetext}
\begin{equation}
H_{2ph}  = \hbar {\omega _0} _{\left( 2 \right)} \sum\limits_i {\left( {b_i^\dag  b_i^{ }  + \frac{1}{2}} \right)}  - \frac{1}{2}V_b \sum\limits_{\left\langle {i,j} \right\rangle } {\left( {b_i^{ }  + b_i^\dag  } \right)\left( {b_j^{ }  + b_j^\dag  } \right)}  + g_2 \sum\limits_{i,\gamma } {\left( {b_i^{ }  + b_i^\dag  } \right)\left( {c_{2i\gamma }^\dag  c_{1i\gamma }^{ }  + c_{1i\gamma }^\dag  c_{2i\gamma }^{ } } \right)}
\label{eq:33}
\end{equation}
\end{widetext}
describes off-diagonal electron-phonon transition processes between the states ``1'' and ``2''. Here,  $g_{1\left( 2 \right)} $ are the constants of electron-phonon interaction, ${\omega _0} _{\left( {1,2} \right)} $ are the frequencies of ``a''- and ``b''-type phonons. The terms proportional to $V_{a\left( b \right)} $ describe interactions of $a$($b$)-phonons at different sites of a crystal lattice.

Within the mean field approximation applied to the matrix Green function 
\begin{eqnarray}
{\mathbf{G}}^\gamma  \left( \omega  \right) &=& \left( {\begin{array}{*{20}c}
   {\left\langle {\left\langle {{c_{1f\gamma } }}
 \mathrel{\left | {\vphantom {{c_{1f\gamma } }^{ } {c_{1g\gamma }^\dag  }}}
 \right. \kern-\nulldelimiterspace}
 {{c_{1g\gamma }^\dag  }} \right\rangle } \right\rangle } & {\left\langle {\left\langle {{c_{1f\gamma } }}
 \mathrel{\left | {\vphantom {{c_{1f\gamma } }^{ } {c_{2g\gamma }^\dag  }}}
 \right. \kern-\nulldelimiterspace}
 {{c_{2g\gamma }^\dag  }} \right\rangle } \right\rangle }  \\
   {\left\langle {\left\langle {{c_{2f\gamma } }^{ }}
 \mathrel{\left | {\vphantom {{c_{2f\gamma } }^{ } {c_{1g\gamma }^\dag  }}}
 \right. \kern-\nulldelimiterspace}
 {{c_{1g\gamma }^\dag  }} \right\rangle } \right\rangle } & {\left\langle {\left\langle {{c_{2f\gamma } }}^{ }
 \mathrel{\left | {\vphantom {{c_{2f\gamma } }^{ } {c_{2g\gamma }^\dag  }}}
 \right. \kern-\nulldelimiterspace}
 {{c_{2g\gamma }^\dag  }} \right\rangle } \right\rangle }  \\
\end{array}} \right)_\omega   \nonumber \\
&=& \left( {\begin{array}{*{20}c}
   {G_{11}^\gamma  } & {G_{12}^\gamma  }  \\
   {G_{21}^\gamma  } & {G_{22}^\gamma  }  \\
\end{array}} \right)_{\omega}
\label{eq:34}
\end{eqnarray}
one has
\begin{eqnarray}
&G&_{11\left(22\right)}^\gamma  \left( \omega  \right) = \frac{{\omega  \mp \frac{\varepsilon }{2} \pm g_1 \Delta _{1ph} }}{{\left( {\omega  - \omega _1 } \right)\left( {\omega  - \omega _2 } \right)}}, \nonumber \\
G_{12}^\gamma  \left( \omega  \right) &=& G_{21}^\gamma  \left( \omega  \right) = \frac{{\left( {z {J_{ex} } \Delta _{ex}  + g_2 \Delta _{2ph} } \right)}}{{\left( {\omega  - \omega _1 } \right)\left( {\omega  - \omega _2 } \right)}}, \nonumber
\end{eqnarray}
where $J_{ex} = J'_{ex} + J''_{ex}$, $\Delta _{1ph} = {\left\langle {a_i^{ }  + a_i^\dag  } \right\rangle }$, $\Delta _{2ph} = {\left\langle {b_i^{ }  + b_i^\dag  } \right\rangle }$, $\Delta _{ex}  = \left\langle {c_{2j\gamma }^\dag  c_{1j\gamma }^{ } } \right\rangle  = \left\langle {c_{1j\gamma }^\dag  c_{2j\gamma }^{ } } \right\rangle $ is the excitonic order parameter, and the dispersion is
\begin{eqnarray}
\omega _{1\left(2\right)}  =  &\pm& \left[ \left( {\frac{\varepsilon }{2} - g_1 \Delta _{1ph} } \right)^2 \right. \nonumber \\
&& \left. + \left( {z J_{ex} \Delta _{ex}  + g_2 \Delta _{2ph} } \right)^2\right]^\frac{1}{2}   =  \pm s.
\label{eq:35}
\end{eqnarray}
In terms of the Matsubara frequencies $\omega _n$ and $\varepsilon _n$, and using the mean field approximation with respect to the phonon-phonon interaction $
V_{a\left( b \right)} $ in the Eqs.~\ref{eq:32}, \ref{eq:33}, one obtains
%\begin{widetext}
\begin{eqnarray}
\Delta _{1ph}  = \left\langle {a_i^{ }  + a_i^\dag  } \right\rangle _{\omega _n  = 0}  = \frac{{2g_1 }}{{\left( {\hbar \omega _{0\left( 1 \right)}  - 2zV_a } \right)}} \nonumber \\
\times\frac{1}{\beta }\sum\limits_{m,\gamma } {\left[ {\left\langle {c_{2i\gamma }^\dag  c_{2i\gamma }^{ } } \right\rangle _{\varepsilon _m }  - \left\langle {c_{1i\gamma }^\dag  c_{1i\gamma }^{ } } \right\rangle _{\varepsilon _m } } \right]}, 
\label{eq:36}
\end{eqnarray}
\begin{eqnarray}
\Delta _{2ph}  \equiv \left\langle {b_i^{ }  + b_i^\dag  } \right\rangle _{\omega _n  = 0}  =  - \frac{{2g_2 }}{{\left( {\hbar \omega _{0\left( 2 \right)}  - 2zV_b } \right)}} \nonumber \\
\times\frac{1}{\beta }\sum\limits_{m,\gamma } {\left[ {\left\langle {c_{1i\gamma }^\dag  c_{2i\gamma }^{ } } \right\rangle _{\varepsilon _m }  + \left\langle {c_{2i\gamma }^\dag  c_{1i\gamma }^{ } } \right\rangle _{\varepsilon _m } } \right]}. 
\label{eq:37}
\end{eqnarray}
%\end{widetext}

Summing over $\varepsilon_n$ in Eqs.~\ref{eq:36}, \ref{eq:37}, one finally obtains
\begin{eqnarray}
\Delta _{1ph}  &=&  - \frac{{4g_1 }}{{\left( {\hbar \omega _{0\left( 1 \right)}  - 2zV_a } \right)}} \nonumber \\
&\times&\frac{{\varepsilon  - 2g_1 \Delta _{1ph} }}{{2s}}\tanh \left( {\frac{s}{{2k_B T}}} \right),
\label{eq:38}
\end{eqnarray}
\begin{eqnarray}
\Delta _{2ph}  &=& \frac{{4g_2 }}{{\left( {\hbar \omega _{0\left( 2 \right)}  - 2zV_b } \right)}} \nonumber \\
&\times&\frac{{\left( {z J_{ex}\Delta _{ex}  + g_2 \Delta _{2ph} } \right)}}{s}\tanh \left( {\frac{s}{{2k_B T}}} \right)
\label{eq:39}
\end{eqnarray}
The expressions for the phonon order parameters given above are valid in the limit when $g_1  <  < \omega _{0\left( 1 \right)}$ and $g_2  <  < \omega _{0\left( 2 \right)}$.

Using the expression
\begin{equation}
\left\langle {c_{\lambda 'g\gamma }^\dag  c_{\lambda f\gamma }^{ } } \right\rangle  =  - \frac{1}{\pi }\int {d\omega f_F \left( {\omega ,\mu } \right){\mathop{\rm Im}\nolimits} G_{\lambda \lambda '}^\gamma  \left( {f - g,\omega  + i\delta } \right)}  \nonumber
\end{equation}
for the correlation function together with Eq.~\ref{eq:34}, the occupation numbers of the states ``1'',``2'' are
\begin{equation}
n_1  = \frac{1}{2}\left[ {1 + \frac{{\varepsilon /2 - g_1 \left\langle {a_i  + a_i^\dag  } \right\rangle }}{s}\tanh \left( {\frac{s}{{2k_B T}}} \right)} \right],
\label{eq:40}
\end{equation}
\begin{equation}
n_2  = \frac{1}{2}\left[ {1 - \frac{{\varepsilon /2 - g_1 \left\langle {a_i  + a_i^\dag  } \right\rangle }}{s}\tanh \left( {\frac{s}{{2k_B T}}} \right)} \right],
\label{eq:41}
\end{equation}
and the excitonic order parameter is
\begin{equation}
\Delta _{ex}  =  - \frac{{\hbar \omega _{0\left( 2 \right)}  - 2zV_b }}{{4g_2 }}\Delta _{2ph}.
\label{eq:42}
\end{equation}

Among the solutions of Eqs.~\ref{eq:38}--\ref{eq:42} we are interested in those, which correspond to the free energy $F =  - k_B T\ln \left( {e^{ - \beta \omega _1 }  + e^{\beta \omega _1 } } \right)$ minimum and satisfy the equation, defining the value of the chemical potential
\begin{eqnarray}
&n&_1  + n_2  = 1 \nonumber \\
&=& -\frac{1}{\pi }\int {d\omega f_F \left( {\omega ,\mu } \right)\left( {{\mathop{\rm Im}\nolimits} G_{11}^\gamma  \left( {\omega  + i\delta } \right) + {\mathop{\rm Im}\nolimits} G_{22}^\gamma  \left( {\omega  + i\delta } \right)} \right)}. \nonumber
\end{eqnarray}

Let us investigate the roles of various interactions in Eq.~\ref{eq:31}. As in Sec.~\ref{sec:4}, we consider the case of a two-dimensional square lattice and neglect $J'_{ex}$. The frequencies $\omega _{0\left( 1 \right)} $ and $\omega _{0\left( 2 \right)} $ in Eqs.~\ref{eq:32}, \ref{eq:33} are supposed to be equal to $0.1J_0$. To simplify the discussion we will also assume that $V_b  = 0$ (an influence of this parameter will be disscussed below). 

\begin{figure}
\includegraphics[width=8.6cm]{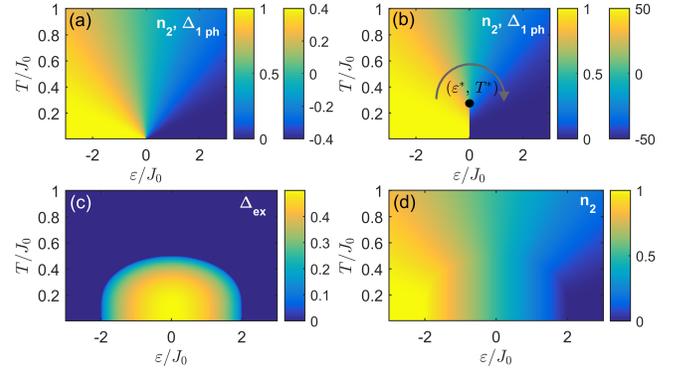}
\caption{\label{fig:6} The mean-field phase diagrams in the following cases: (a) $g_1  = 0.01J_0 $, $J''_{ex}  = 0$, $V_a  = 0$; (b) $g_1  = 0.01J_0 $, $J''_{ex}  = 0$, $V_a  = 0.0124J_0 $, (c), (d) $J''_{ex}  =  - 0.5J_0 $, $V_a  = 0$, $g_1 = 0$.  In (a) and (b), the first color scale corresponds to $n_2$, the second corresponds to $\Delta _{1ph} $.}
\end{figure}

Firstly, we consider the case $g_2  = 0$. In Fig.~\ref{fig:6} (a),(b) the phase diagrams of the phonon order parameter $\Delta _{1ph} $ (Eq.~\ref{eq:38}) and the occupation number $n_2$ (Eq.~\ref{eq:41}) are shown at $J''_{ex}  = 0$, $g_1  = 0.01J_0 $ in the case when $V_a  = 0$ (a) and $V_a  = 0.0124J_0 $ (b). In the first case (a) there is a smooth crossover, whereas in the second (b) there is an isostructural first-order phase transition with the critical ``vapor-liquid'' point $\left(\varepsilon ^ *  ,T^ *\right)$ present - above this point, the system can be smoothly transformed from a state with $n_2=1$ to a state with $n_2=0$. The case $V_a  \ne 0$ and $g_1  = 0$ is trivial, since, according to Eq.~\ref{eq:38}, $\Delta _{1ph}  = 0$, and the Hamiltonian Eq.~\ref{eq:32} can be diagonalized by a canonical transformation as $H_{1ph}  = \sum\limits_{\mathbf{q}} {\omega \left( \mathbf{q} \right)\left( a_\mathbf{q}^{\dag}  a_\mathbf{q}^{ }  + 1/2 \right)} $, where $\omega \left( \mathbf{q} \right) = \hbar \omega _{0\left( 1 \right)} \sqrt {\left( {1 - \frac{{4V_a }}{{\hbar \omega _{0\left( 1 \right)} }}\left( {\cos q_x  + \cos q_y } \right)} \right)} $, which describes an ideal gas of phonons, in which, as in Fig.~\ref{fig:6}, there is no phase transition. Comparing Fig.~\ref{fig:6} (a) and (b), it can be seen that the intersite interaction $V_a$ provides the cooperativity necessary for a phase transition. Each of the interactions $V_a$ and $g_1$ separately does not lead to a phase transition (the presence of one of them separately is neccessary, but not sufficient). In the following we will exclude the phonon mechanism of cooperativity by assuming $V_a=0$.

Whereas the phase diagram shown in Fig.~\ref{fig:6} (a) corresponds to the quantum phase transition at $T=0$ in the absence of cooperativity, in Fig.~\ref{fig:6} (b) there is a first order phase transition at finite temperatures up to the critical point $T^* $. In Fig.~\ref{fig:6} (c) and (d) the phase diagram of the excitonic order parameter $\Delta _{ex} $ given by Eq.~\ref{eq:42} and the phase diagram of the occupation number $n_2$ given by Eq.~\ref{eq:41} are shown at $J''_{ex}  =  - 0.5J_0 $ when $V_a  = 0$ and $g_1  = 0$. In such a case there exists the second order phase transition into the excitonic condensate phase (see Fig.~\ref{fig:6} (c)).

\begin{figure}
\includegraphics[width=8.6cm]{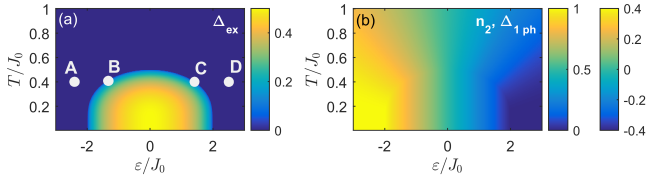}
\caption{\label{fig:7} The phase diagrams of (a) the excitonic order parameter $\Delta_{ex}$, (b) the occupation number $n_2$ of the one-electron state ``2'' and the phonon order parameter $\Delta _{1ph}$ at $g_2=0$ and $V_a=0$. In (b), the first color scale corresponds to $n_2$, the second corresponds to $\Delta _{1ph}$}
\end{figure}

The results of calculations in the case when $\Delta _{ex} $, $n_2$ and $\Delta _{1ph} $ are nonzero at the same time, but $V_a$ is zero, are shown in Fig.~\ref{fig:7} (a), (b) at $J''_{ex}  =  - 0.5J_0 $ and $g_1  = 0.01J_0 $. From Figs.~\ref{fig:6} (a), (b) and \ref{fig:7} (b) it is clear that $n_2$ and $\Delta _{1ph} $ behave similarly (see Eqs.~\ref{eq:38}, \ref{eq:41}).

Analogously to Eq.~\ref{eq:18} one obtains the two-electron Green function

\begin{eqnarray}
\left\langle \left\langle c_{1f\gamma }^\dag  c_{2f\gamma }^{ } | c_{2g\gamma }^\dag  c_{1g\gamma }^{ } \right\rangle\right\rangle = \frac{\left( {n_1  - n_2 } \right)} {\left( {\omega  - \omega _ +  } \right)\left( {\omega  - \omega _ -  } \right)} \nonumber \\
\times
\left[ \omega  + \varepsilon  - 2g_1 \Delta_{1ph}  + \left( {n_1  - n_2 } \right)J'_{ex} \left( \mathbf{k} \right) \right],
\label{eq:43}
\end{eqnarray}
\begin{eqnarray}
\omega_{\pm}  \left( \mathbf{k} \right) =  &\pm& \left[ \left( {\varepsilon  - 2g_1 \Delta_{1ph}  + \left( {n_1  - n_2 } \right){J'}_{ex} \left( \mathbf{k} \right)} \right)^2 \right. \nonumber \\ 
&-& \left. \left( {n_1  - n_2 } \right)^2 {J''}_{ex}^2 \left( \mathbf{k} \right) \right]^{\frac{1}{2}}
\label{eq:44}
\end{eqnarray}

\begin{figure}
\includegraphics[width=8.6cm]{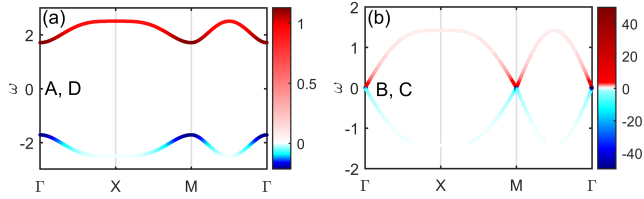}
\caption{\label{fig:8} The spectrum of excitons $\omega _ \pm  \left( \mathbf{k} \right)$ given by Eq.~\ref{eq:44} calculated at the points $A$, $B$, $C$, and $D$ of the phase diagram Fig.~\ref{fig:7} (a) when $g_2=0$.}
\end{figure}

The spectrum of excitons given by Eq.~\ref{eq:44} at the points $A$, $B$, $C$, and $D$ of the phase diagram Fig.~\ref{fig:7} (a) is shown in Fig.~\ref{fig:8}. We define the excitonic spectrum gap as $E_g  = \omega _ +  \left( \mathbf{k} \right) - \omega _ -  \left( \mathbf{k} \right)$, where $\mathbf{k} = \Gamma \left(\text{M}\right)$. The diagonal electron-phonon interaction does not lower the symmetry of the Hamiltonian in Eq.~\ref{eq:30}, so the gap $E_g$ is zero at the boundary of the excitonic condensate phase (see Fig.~\ref{fig:8}, $B$ and $C$).

\begin{figure}
\includegraphics[width=8.6cm]{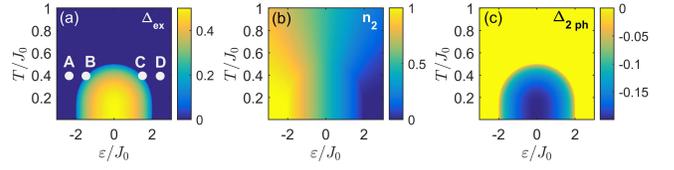}
\caption{\label{fig:9} The phase diagrams of the (a) excitonic order parameter $\Delta_{ex}$, (b) occupancy $n_2$, and (c) phonon order parameter $\Delta_{2ph}$ in the case $g_1=0$.}
\end{figure}

\begin{figure}
\includegraphics[width=8.6cm]{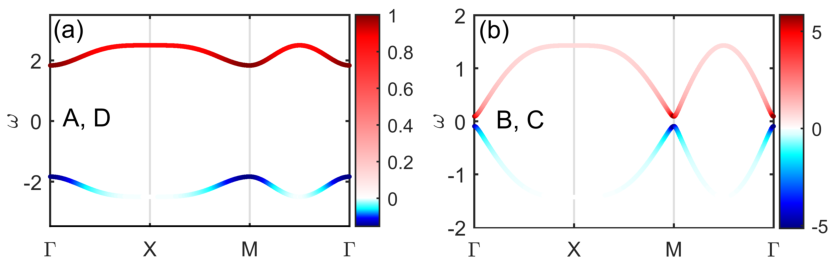}
\caption{\label{fig:10} The spectrum of excitons $\omega _ \pm  \left( \mathbf{k} \right)$ given by Eq.~\ref{eq:44} calculated at the points $A$, $B$, $C$, and $D$ of Fig.~\ref{fig:9} (a) when $g_2\neq0$.}
\end{figure}

Let us now discuss the case $g_1 = 0$, but $g_2 = 0.01J_0$ ($J''_{ex}  =  - 0.5J_0 $ as before) shown in Fig.~\ref{fig:9}. Due to Eq.~\ref{eq:42} the behavior of $\Delta _{ex} $ and $\Delta _{2ph}$ is qualitatively identical. In this case, the non-diagonal electron-phonon interaction lowers the symmetry of the Hamiltonian in Eq.~\ref{eq:30}, thus, the gap $E_g$ is finite and remains open at the excitonic condensate boundary (see Fig.~\ref{fig:10}, points $B$ and $C$).

As one can see from Figs.~\ref{fig:8} and \ref{fig:10} (similarly to Figs.~\ref{fig:3}--\ref{fig:5}), the spectral weight is distributed non-uniformly among the Brillouin zone. We note that since at $\varepsilon = 0$ a change of the ground state takes place, the system cannot be smoothly (adiabatically) transformed from the state at the point $A$ to the state at the point $D$. This way, across the line $\varepsilon = 0$, as one can see from Eq.~\ref{eq:44} ($n_1=n_2=0.5$ at $\varepsilon = 0$), the gap becomes zero even when the non-diagonal electron-phonon interaction is present. 

\section{\label{sec:6} Discussion and conclusions}

Using Eq.~\ref{eq:1}, one can consider two limiting cases. In the first (weakly correlated), when $H_{Coulomb}  << H_\Delta + H_t$, one deals with a two-band semiconductor or a semimetal, depending on the $t/\Delta$ ratio. In this case, the Bose-Einstein or BCS formation of excitonic condensate is possible. In the second case (strongly correlated), when the energy of the Coulomb interaction of electrons is comparable to the crystal field energy $H_{Coulomb}  \sim H_\Delta$ and larger than their kinetic energy $H_{Coulomb}  > H_t$, there appears a possibility for the spin crossover and formation of localized magnetic excitons. In the present paper we have shown in the framework of the two-band Hubbard-Kanamori model that the condensation of such excitons takes place near the spin crossover and leads to the appearance of the antiferromagnetic ordering even when an interatonic exchange interaction is absent. In other words, the appearance of magnetism caused by excitonic condensation is found. It should be noted that in the exciton dielectric model at weak electron-electron interaction, the formation of excitonic condensate can lead to magnetic ordering without exchange interaction in a similar way \cite{Volkov}.

Close to the phase transition point the elementary excitation spectra are known for their significant dependency on the type of statistics: whereas in the fermion system, a gapped branch with a gap width proportional to the order parameter exists as well as a gapless branch, in the Bose system only the latter exists. This is due to the fact that fermionic systems have both individual gapped and collective gapless excitations, while Bose systems can only have collective excitations (as was shown using the diagrammatic approach in Ref.~\onlinecite{Migdal}).

In the present paper, we have obtained the spectrum of exciton excitations and shown its instability towards the formation of the excitonic Bose-condensate. This spectrum describes collective, from the point of view of the electronic (Fermi) system, excitations, but one can interpret this spectrum as a quasiparticle (one-particle) one with respect to Bose-type particles described by the Hamiltonian in Eq.~\ref{eq:13}. Everywhere outside the excitonic condensate phase there is a gap in the spectrum, which becomes zero at the boundary of the second order phase transition, which agrees with the general idea, that, below the point of a phase transition, there should arise a gapless Goldstone mode, describing collective excitations in the excitonic condensate phase \cite{Nasu, Murakami}. In other words, the appearance of such a gapless mode is preceded by the closing of the gap in the quasiparticle excitations spectrum. The non-diagonal electron-phonon interaction (contrary to the diagonal one) leads to the opening of a gap in the individual excitonic excitations spectrum at the boundary of the second order phase transition, which is consistent with the result of the work \onlinecite{Murakami}, where it was shown that the collective Goldstone mode acquires mass due to non-diagonal electron-phonon interaction. In this case, the Bose spectrum of excitations (one-particle excitonic and collective in the excitonic phase) has a gap on both sides of the phase transition. According to the authors of the work \onlinecite{Murakami}, this circumstance plays an important role in the photoinduction of an excitonic condensate and can provide a new strategy of enhancing the order parameter in analogous systems (such as superconductors). An interesting feature of the spectra that we obtained is the nonuniform spectral weight distribution among the Brillouin zone. In this connection, it is interesting to investigate the behavior of collective excitations in the exciton condensate phase. 

\begin{acknowledgments}
The reported study was funded by the Russian Scientific Fund, grant No 18‐12‐00022.
\end{acknowledgments}

\bibliography{paper}

\end{document}